\documentclass[10pt,twocolumn,letterpaper]{article}

\usepackage[pagenumbers]{cvpr} 

\usepackage[dvipsnames]{xcolor}

\definecolor{cvprblue}{rgb}{0.21,0.49,0.74}
\usepackage[pagebackref,breaklinks,colorlinks,citecolor=cvprblue]{hyperref}
\usepackage{multirow}
\usepackage{tabularx,booktabs}
\newcolumntype{C}{>{\centering\arraybackslash}X} 
\usepackage[dvipsnames]{xcolor}
\usepackage{longtable}
\usepackage{makecell, cellspace, caption}
\usepackage{array}
\usepackage{subcaption}
\usepackage{cuted}
\usepackage{multirow}
\usepackage[utf8]{inputenc}
\usepackage{tabularx}
\usepackage{array}
\usepackage{amssymb}
\usepackage{makecell}
\usepackage[margin=2cm]{geometry}
\usepackage{adjustbox}
\usepackage{multirow}
\usepackage{tabulary,booktabs}
\usepackage{ragged2e}
\usepackage{amsmath}
\usepackage{booktabs}
\usepackage{array}
\newcolumntype{L}{>{$}l<{$}}
\newcolumntype{C}{>{$}c<{$}}
\newcolumntype{R}{>{$}r<{$}}

\usepackage{amssymb}
\usepackage{pifont}
\usepackage{algorithm2e}
\RestyleAlgo{ruled}
\SetKw{Kwparam}{Params:}
\SetKw{Kwensure}{Ensure:}
\SetKwComment{Comment}{/* }{ */}

\def\paperID{9153} 
\def\confName{CVPR}
\def\confYear{2024}

\title{LATIS: Lambda Abstraction-based Thermal Image Super-resolution}

\author{Gargi Panda$^1$, Soumitra Kundu$^1$, Saumik Bhattacharya$^1$, Aurobinda Routray$^1$\\
$^1$IIT Kharagpur\\
{\tt\small pandagargi@gmail.com, soumitra2012.kbc@gmail.com, saumik@ece.iitkgp.ac.in, aroutray@iitkgp.ac.in}
}

\begin{document}
\maketitle
\begin{abstract}
Single image super-resolution (SISR) is an effective technique to improve the quality of low-resolution thermal images. Recently, transformer-based methods have achieved significant performance in SISR. However, in the SR task, only a small number of pixels are involved in the transformer's self-attention (SA) mechanism due to the computational complexity of the attention mechanism.  The lambda abstraction is a promising alternative to SA in modeling long-range interactions while being computationally more efficient. This paper presents lambda abstraction-based thermal image super-resolution (LATIS), a novel lightweight architecture for SISR of thermal images. LATIS sequentially captures local and global information using the local and global feature block (LGFB). In LGFB, we introduce a global feature extraction (GFE) module based on the lambda abstraction mechanism, channel-shuffle and convolution (CSConv) layer to encode local context. Besides, to improve the performance further, we propose a differentiable patch-wise histogram-based loss function. Experimental results demonstrate that our LATIS, with the least model parameters and complexity, achieves better or comparable performance with state-of-the-art methods across multiple datasets.  
\\

\end{abstract}    
\section{Introduction}
\label{sec:intro}
Thermal images, unfazed by extreme lighting conditions unlike visible images, play a pivotal role in applications such as disease diagnosis \cite{disease1,disease2}, military and surveillance \cite{survei}, pedestrian tracking, and automated driving \cite{vehi}. Still, consumer-oriented thermal cameras typically exhibit low resolution due to hardware limitations \cite{infrared, icvs_thermal}. Thermal image super-resolution (TISR) addresses this limitation by enhancing the resolution of thermal images in order to improve their utility in respective
applications.
\begin{figure}[t]
  \centering
     \includegraphics[width=0.5\textwidth]{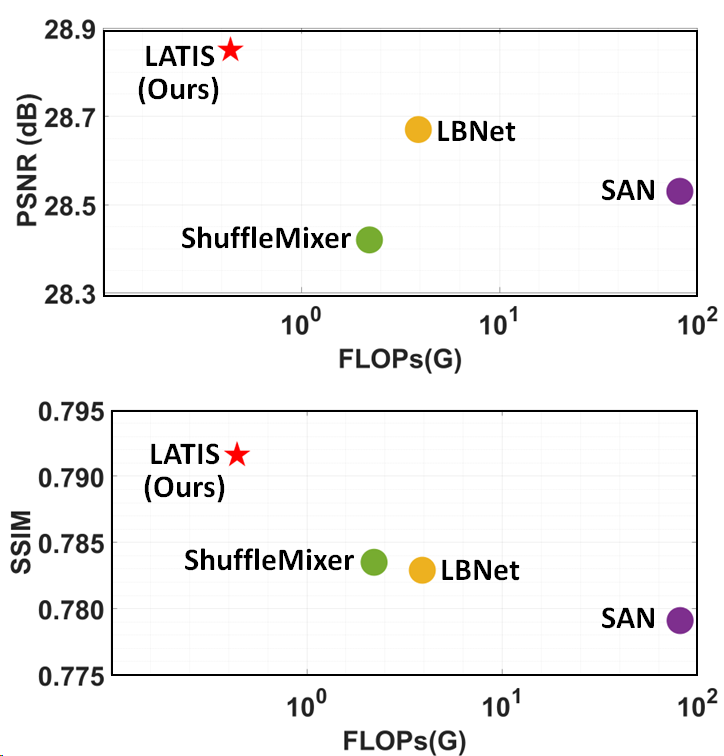}

   \caption{Performance comparison of PSNR/SSIM and model complexity on FLIR-ADAS \cite{flir_adas} test set for $\times$4 SR. LATIS, with the least model complexity, surpasses the state-of-the-art methods SAN \cite{san}, ShuffleMixer \cite{shuffleMixer} and LBNet \cite{lbnet}.}
   \label{fig:flops}
\end{figure}

Most often, thermal cameras come embedded with an RGB camera, which paves the way for guided super-resolution (GSR) methods employing both the thermal and RGB images \cite{ista, ugsr}. However, guided techniques may encounter the challenge of over-
transferring RGB textures. 

Contrary to GSR methods, single image super-resolution (SISR) reconstructs a high-resolution (HR) image solely from a low-resolution (LR) one. Following SRCNN \cite{srcnn}, convolutional neural network (CNN)-based SISR methods have significantly improved with time \cite{pixelshuffle,rdn,fsrcnn,srgan,esrgan,lapar,shuffleMixer}. Convolution operation is very efficient in extracting the local features, but it shows limited performance while capturing long-range dependencies. Recently, there has been a significant rise in applying transformer-based models such as vision
transformer (ViT) \cite{vit} for low-level vision tasks \cite{swinir,uformer,oneformer}, including SR \cite{hat,lbnet,dat}. The self-attention (SA) \cite{attention} module in transformer seems very promising in modeling long-range dependencies. However, the space and computational complexity of SA increases quadratically with the input size, making it infeasible to apply to high-resolution feature maps in SR tasks. To reduce the complexity in low-level vision tasks, ViT-based models \cite{dat, hat, uformer, swinir} divide the input feature map into non-overlapping windows and compute SA on each window independently. This limits the number of pixels engaged in SA. 

As an alternative to SA, recently proposed lambda networks \cite{lambdanetworks} use the lambda layer to capture long-range dependencies. Instead of computing attention map like SA, the lambda layer uses content and position-based linear functions, termed  as lambda abstraction. This lambda abstraction based modeling of long-range interactions makes the lambda layer computationally more efficient than SA. As a result, the lambda layer can be applied to high-resolution feature maps, which enables them to model interactions between all the pixels in the feature map to provide global dependency. While this lambda abstraction-based modeling of global dependency is successfully applied in image recognition, its efficacy in SR tasks still lacks exploration.  

In this paper, we introduce lambda abstraction-based thermal image super-resolution (LATIS), a unique network design for thermal SR. Our LATIS sequentially captures local and global features with local and global feature block (LGFB). Specifically, we propose a global feature extraction (GFE) module in LGFB to model global dependency based on the lambda abstraction mechanism. Besides, we develop a channel shuffle and convolution (CSConv) layer to extract local information. Our ablations show that these design choices allow LATIS to encode information efficiently for better SR.

Moreover, we formulate an image histogram-based loss function to further improve the quality of TISR.  Matching the histogram between two images can constrain their similarity of color distribution \cite{color, probability, huenet}. 
This motivates us to explore the effectiveness of a differentiable histogram-based loss function in SR task. Instead of comparing the histogram of the full image \cite{huenet}, we adopt a patchwise histogram comparison to maintain structural similarity between the image pairs. We use the earth mover’s distance (EMD) between the histograms for constructing our novel patchwise EMD loss. 

LATIS, designed with LGFB as a core module and trained with the $L_1$ and proposed patch-wise EMD loss, obtains state-of-the-art performance while maintaining minimal computation complexity. Figure \ref{fig:flops} shows the performance comparison of LATIS with three recent SISR methods on FLIR-ADAS \cite{flir_adas} test set for x4 SR. LATIS achieves the highest PSNR and SSIM with the least FLOPs. 
Our contributions can be summarized as:
\begin{enumerate}
    \item We introduce LATIS, a lightweight SISR network for thermal images. Specifically, we develop a GFE module to capture global dependency based on the lambda abstraction mechanism and a CSConv layer to explore the local context. Furthermore, we propose an LGFB to encode local and global features sequentially. 
    \item We propose a differentiable patch-wise histogram-based loss function that uses Earth mover's distance to compare two histograms.
    \item Experimental results show that our algorithm achieves state-of-the-art performance with the least model complexity and parameters.
\end{enumerate}

In the rest of this paper, Section \ref{sec:litReview} summarizes the related work with image super-resolution and image histogram-based loss function. Section \ref{sec:method} describes the proposed LATIS architecture and patch-wise EMD loss. In Section \ref{sec:experiments}, we discuss the experimental analysis, including training settings, dataset details, quantitative and qualitative comparison, and ablation study. We conclude the work in Section \ref{sec:conclusion}.
\section{Related Work}
\label{sec:litReview}
In thermal SR, both GSR and SISR techniques are used. The most common application of GSR is depth map super-resolution \cite{msgnet, depthsrnet}. The recent depth SR method, DCTNet \cite{dctnet}, extracts cross-modal information between depth and RGB images using discrete cosine transform and uses it in reconstructing the depth HR map. MMSR \cite{mmsr} proposes a mutual modulation strategy between depth and RGB images in depth SR. Recently, GSR methods are also used for thermal images \cite{ista, ugsr}. Gupta et al. \cite{ugsr} used input\:(UGSR-ME) or feature\:(UGSR-FA) based alignment loss to improve the thermal SR performance. But, GSR methods may
encounter the problem of over-transferring RGB textures especially while performing thermal SR.

Many CNN-based methods \cite{srcnn,vdsr,edsr,srgan,esrgan,san, lapsrn,lapar,shuffleMixer} have achieved significant performance in SISR task. SAN \cite{san} considers the interdependencies between features using second-order channel attention for improved SR. Shuffle mixer \cite{shuffleMixer} uses large kernel with channel shuffle and split operations to improve the performance. However, convolution operation has a local receptive field, which limits the performance of CNN in modeling global dependency.

Transformer \cite{attention}, with the self-attention mechanism, is effective in extracting long-range information. Dosovitskiy et al. \cite{vit} used vision transformer (ViT) for image classification; since then, many researchers have proposed vanilla ViT-based models for low-level vision tasks \cite{swinir,uformer,lbnet,dat}. LBNet \cite{lbnet} uses a recursive transformer-based model for SISR. However, the self-attention mechanism has quadratic complexity with the input size, which restricts its application to high-resolution feature maps. Due to this limitation, ViT based methods \cite{hat,lbnet,dat} divide the feature map into non-overlapping local windows, and SA is applied to each window independently. Such an approach limits the scope of SA.

To combat this complexity limitation of SA, linear attention mechanisms \cite{linformer, performer} and lambda layer \cite{lambdanetworks} are promising alternatives. Instead of constructing an attention map like SA, the lambda layer uses content and position-based linear functions, termed as lambda abstraction. This lambda abstraction based method is computationally more efficient than SA in capturing long-range interactions. In our LATIS network, we design the GFE layer to model global dependency based on the lambda abstraction mechanism.

\begin{figure*}[hbt!]
    \centering
     \includegraphics[width=\textwidth]{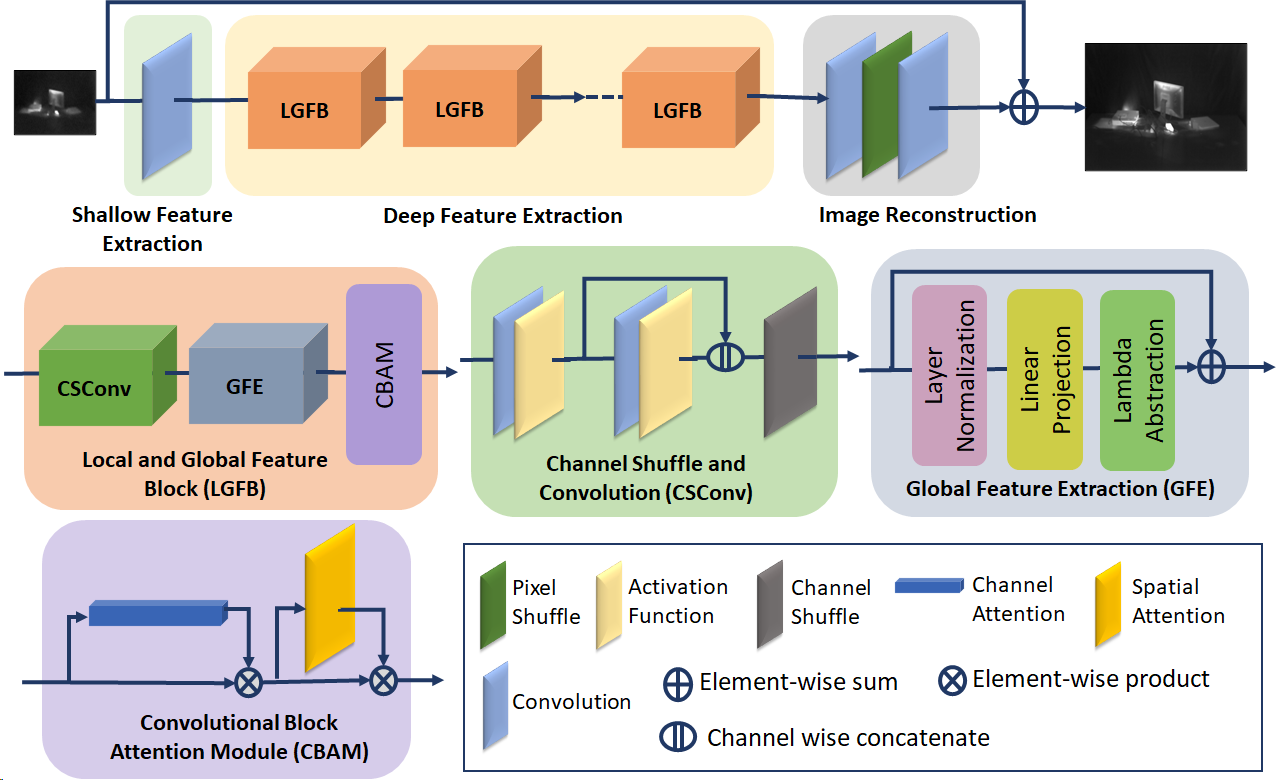}
      \caption{An overview of LATIS and the structure of LGFB}
      \label{fig:net}
\end{figure*}

Image histogram-based loss is mainly used in image-to-image translation task \cite{color, probability, huenet}.
Classical methods for color transfer \cite{color, probability} use histogram matching between a given and a target image. However, histogram operation in its standard form is non-differentiable, which restricts its application in loss function in gradient-based optimization of deep neural networks. Aharon et al. \cite{huenet} proposed a method for constructing a smooth and differentiable histogram and used image histogram similarity loss in an image-to-image translation task. However, they considered histogram of the full image for formulating the loss function. This global comparison of color distribution ignores the images' local structural similarity. In our patchwise EMD loss, we propose a differentiable loss function that compares the image pairs based on patch-wise histogram similarity.
\section{Proposed Method}
\label{sec:method}
In this section, we first describe the overall architecture of LATIS, a lightweight network for thermal image super-resolution. Then, we detail LGFB, the core module of LATIS. Finally, we formulate our histogram-based patch-wise EMD loss. 

\subsection{LATIS Architecture}
Figure \ref{fig:net} shows an overview of LATIS. The low-resolution single channel input thermal image, $I_{LR}\in\mathbb{R}^{H,W,1}$, is processed in three steps: shallow feature extraction, deep feature extraction, and image reconstruction.  We apply a convolution layer for generating the shallow feature $F_S\in\mathbb{R}^{H,W,C}$, where $H\times W$ denotes the resolution of the input image, and C is the number of feature channels. Following the shallow feature extraction step, a series of LGFBs are employed to acquire deep feature $F_{D}\in\mathbb{R}^{H, W, C}$.    

From the deep feature $F_{D}$ with sufficient semantic information, we reconstruct the high-resolution output image $I_{SR}\in\mathbb{R}^{H_{out},W_{out},1}$ using the pixel shuffle method \cite{pixelshuffle}. Given deep feature $F_{D}\in\mathbb{R}^{H,W,C}$, first a point-wise convolution layer is applied to generate intermediate feature $F_{C}\in\mathbb{R}^{H,W,s^2C}$, where $s$ is the scale factor. Then, the pixel shuffle layer rearranges the elements of $F_C$ to form an upsampled feature map $F_{P}\in\mathbb{R}^{sH,sW,C}$. For scale factor $\times$2 and $\times$3, we apply this pixel shuffle-based method once. For scale factor $\times$4, we progressively upsample by repeating this method twice, taking $s=2$. At the end of upsampling, one convolution layer is applied to aggregate the features. Finally, a bicubic upsampled version of the input image $I_{LR}$ is added to the reconstructed image as a residual connection. In the supplementary material, we conduct experiments to show the effectiveness of this residual connection.
\begin{figure*}[hbt!]
    \centering
     \includegraphics[width=\textwidth]{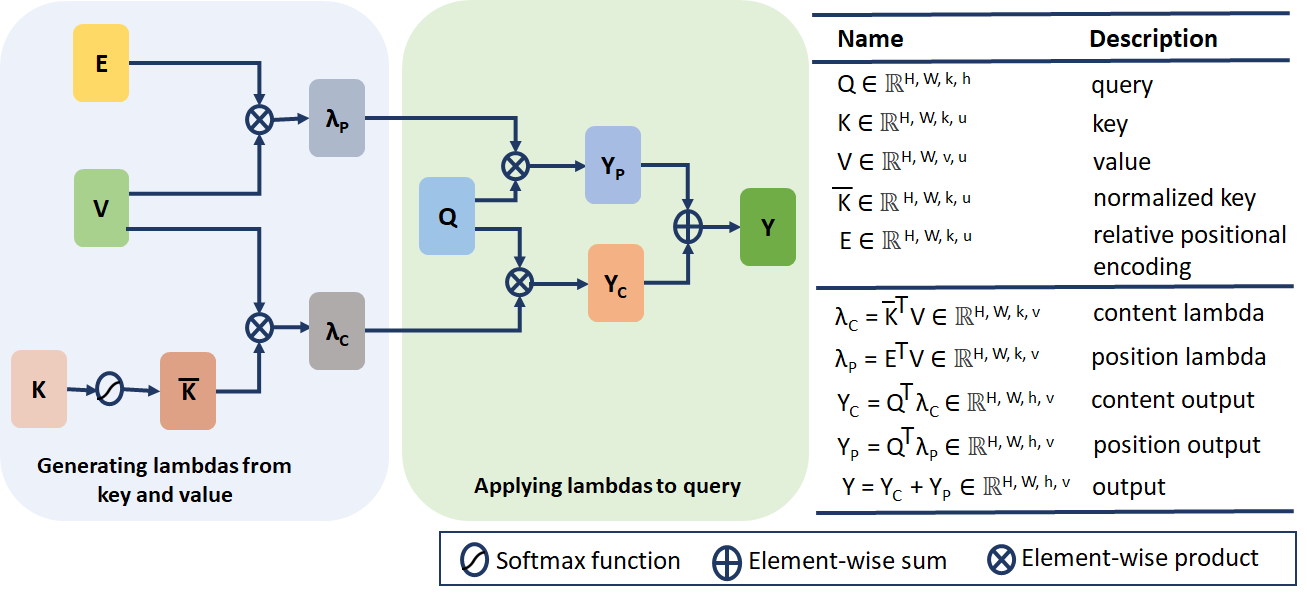}
      \caption{Computational graph of the lambda abstraction mechanism}
      \label{fig:lambda}
\end{figure*}
\subsection{Local and Global Feature Block (LGFB)}
The LGFB is designed to sequentially explore local and global features for efficiently encoding information. In each LGFB, the input feature undergoes three stages: a CSConv layer to encode local context $F_l$, a GFE layer to extract global feature $F_G$, and subsequently, a convolutional block attention module (CBAM) \cite{cbam} for refining the feature map. 

CBAM is a simple and effective module for adaptive feature refinement. It sequentially calculates attention maps from the feature along the channel and spatial dimensions. These attention maps are multiplied with the input feature to highlight more distinctive information.
\subsubsection{Channel Shuffle and Convolution (CSConv) layer}
CSConv layer is developed using convolution and channel shuffle operations to extract local feature $F_l\in\mathbb{R}^{H,W,C}$. Given the shallow feature $F_S\in\mathbb{R}^{H,W,C}$, we sequentially apply two convolution layers with different kernel sizes to learn low-level features at different scales. Then these multi-scale features are aggregated using concatenation.  Following this, channel shuffling operation is employed to exchange information. This procedure can be formulated as follows,
\begin{align} 
&F_{l1}=\sigma(W(F_S))\in\mathbb{R}^{H,W,\frac{C}{2}}\\ 
&F_{l2}=\sigma(W(F_{l1}))\in\mathbb{R}^{H,W,\frac{C}{2}}\\
&F_{l}=\text{Shuffle}(\text{Concat}([F_{l1},F_{l2}])) \in\mathbb{R}^{H,W,C}
\end{align}
where $\sigma$ is the SiLU activation, $W$ is the convolution operation, Concat($\cdot$) and Shuffle($\cdot$) represent the concatenation and shuffling of feature in the channel dimension.  

\begin{figure*}[hbt!]
    \centering
    \begin{subfigure}[t]{0.19\textwidth}
       \includegraphics[width=\textwidth]{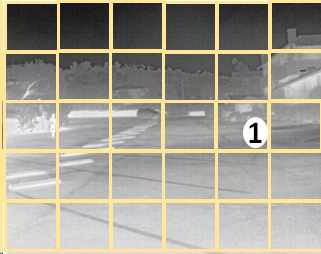}
        \caption{Splitting image into non-overlapping patches}
        \label{fig:gull}
    \end{subfigure}
    \hfill
    \begin{subfigure}[t]{0.28\textwidth}
       \includegraphics[width=\textwidth]{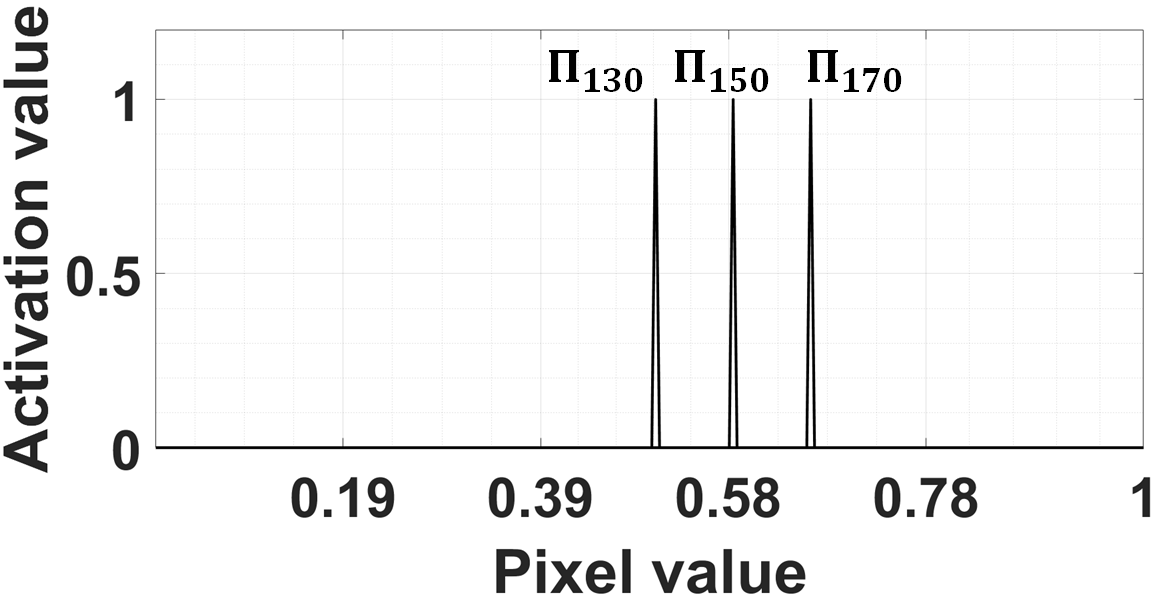}
        \caption{Three out of 256 activation functions, for $k$\:=\:$130$,\:$150$,\:and\:$170$ }
        \label{fig:gull}
    \end{subfigure}
    \hfill
    \begin{subfigure}[t]{0.48\textwidth}
       \includegraphics[width=\textwidth]{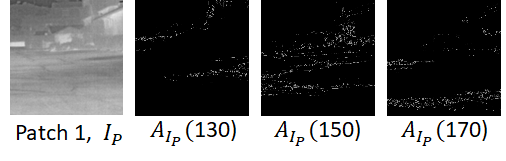}
        \caption{Three out of 256 activation maps of Patch 1}
        \label{fig:gull}
    \end{subfigure} 
    \caption{Differentiable histogram construction using activation functions. (a) We split the image into non-overlapping patches. (b) Activation functions that are smooth and differentiable (c) Applying 256 activation functions to an image patch generates 256 activation maps. Pixels with values closer to $\frac{k}{255}$ have higher values in the k-th activation map. Summation of pixel values in k-th activation map gives the value of k-th bin in the histogram.}
    \label{fig:hist}
\end{figure*}
\subsubsection{Global Feature Extraction (GFE)}
\label{gfe}
The GFE layer is designed based on the lambda abstraction mechanism to capture interactions between all the pixels in the input feature map to provide global dependency. Given the local feature map $F_l\in\mathbb{R}^{H,W,C}$, first, we perform layer normalization to normalize the distribution of intermediate layers \cite{layerNorm}. 

Following this, pointwise convolution is applied in the linear projection layer to generate the query $\big(Q\in \mathbb{R}^{H,W,k\times h}\big)$, key $\big(K\in \mathbb{R}^{H,W,k\times u}\big)$, and value $\big(V\in \mathbb{R}^{H,W,v\times u}\big)$ matrices. The notation $k$ is query/key depth, and $v$ is value depth. We then split query into h heads, and key and value into u heads: $Q\in \mathbb{R}^{H,W,k,h}$, $K\in \mathbb{R}^{H,W,k,u}$, $V\in \mathbb{R}^{H,W,v,u}$. 
  
We apply the lambda abstraction mechanism to these query, key, and value matrices to model global dependency in the input. Figure \ref{fig:lambda} illustrates the implementation of the lambda abstraction mechanism. Lambda abstraction refers to content and position based linear functions, which consider the content and position based interactions in the input feature, respectively.

First, key and value matrices generate content and position lambda functions. For generating content lambda $\lambda_C$, the key is normalized using the softmax operation and then multiplied with value. Positional lambda $\lambda_P$ is computed from the value matrix using relative positional encoding. We use 3D convolution operation to encode the position based interactions in a local neighborhood. Such convolution based positional encoding is effective in SR task. The content and position lambdas can be denoted as,
\begin{equation}
\lambda_C = \overline{K}^TV, \:
    \lambda_P = W_{3D}(V)
\end{equation}
where $\overline{K}$ denotes the normalized key and $W_{3D}(\cdot)$ is the 3D convolution operation.
These content and position lambdas are then applied to the query to generate content output $Y_C$ and position output $Y_P$, respectively. The output $Y$ is the summation of content output and position output. The process can be summarized as,
\begin{equation}
Y_C = Q^T\lambda _C ,\:
Y_P = Q^T\lambda _P,\:
Y=Y_C+Y_P
\end{equation}
Finally, $h$ heads in $Y$ are merged together to get global feature $F_G \in \mathbb{R}^{H,W,C}$.

\subsection{Loss function}
\label{sec:loss}
During training LATIS, $L_1$ loss is used as the content loss to minimize the difference between the SR and HR images,
\begin{equation}
    \mathcal{L}_C = ||\:I_{SR}-I_{HR}\:||_1
\end{equation}
 where  $I_{SR}$ and $I_{HR}$ are the super-resolved image and ground truth image, respectively.
Along with $\mathcal{L}_C$, we use our proposed patchwise EMD loss, $\mathcal{L}_P$ in the initial few epochs. $\mathcal{L}_P$ is detailed in the section below. The content loss and patch-wise EMD loss form the complete loss function,
\begin{equation}
    \mathcal{L} = \mathcal{L}_C+\lambda\mathcal{L}_P
\label{eq:lossfn}
\end{equation}
where,
\[
\lambda = 
\begin{cases}
  \:0.125 & \text{\:if\:no.\:of\:epochs}<\text{n} \\
  \:0 & \text{otherwise}
\end{cases}
\]
The values of $\lambda$ and $n$ are set empirically. In the supplementary material, we conduct experiments for different values of $\lambda$ and $n$.
\begin{figure*}[hbt!]
    \centering
    \begin{subfigure}[t]{0.195\textwidth}
       \includegraphics[width=\textwidth]{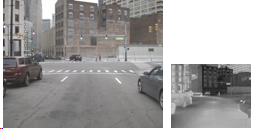}
        \caption{RGB/ Thermal LR}
        \label{fig:gull}
    \end{subfigure}
    \hfill
    \begin{subfigure}[t]{0.195\textwidth}
       \includegraphics[width=\textwidth]{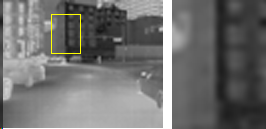}
        \caption{Bicubic}
        \label{fig:gull}
    \end{subfigure}
    \hfill
    \begin{subfigure}[t]{0.195\textwidth}
       \includegraphics[width=\textwidth]{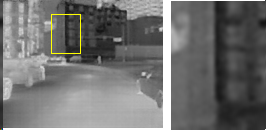}
        \caption{SAN \cite{san}}
        \label{fig:gull}
    \end{subfigure}
    \hfill
    \begin{subfigure}[t]{0.195\textwidth}
       \includegraphics[width=\textwidth]{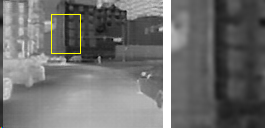}
        \caption{ShuffleMixer \cite{shuffleMixer}}
        \label{fig:gull}
    \end{subfigure}    
    \hfill    
    \begin{subfigure}[t]{0.195\textwidth}
       \includegraphics[width=\textwidth]{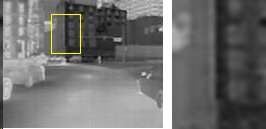}
        \caption{LBNet \cite{lbnet}}
        \label{fig:gull}
    \end{subfigure}
    \hfill
    \begin{subfigure}[t]{0.195\textwidth}
       \includegraphics[width=\textwidth]{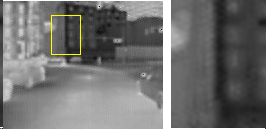}
        \caption{MMSR \cite{mmsr}}
        \label{fig:gull}
    \end{subfigure}
    \hfill    
    \begin{subfigure}[t]{0.195\textwidth}
       \includegraphics[width=\textwidth]{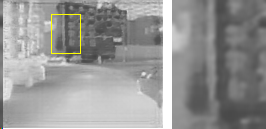}
        \caption{UGSR-FA \cite{ugsr}}
        \label{fig:gull}
    \end{subfigure}
    \hfill
    \begin{subfigure}[t]{0.195\textwidth}
       \includegraphics[width=\textwidth]{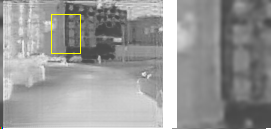}
        \caption{UGSR-ME \cite{ugsr}}
        \label{fig:gull}
    \end{subfigure}
    \hfill
    \begin{subfigure}[t]{0.195\textwidth}
       \includegraphics[width=\textwidth]{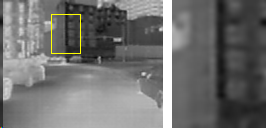}
        \caption{LATIS (\textbf{Ours})}
        \label{fig:gull}
    \end{subfigure}
    \hfill
    \begin{subfigure}[t]{0.195\textwidth}
       \includegraphics[width=\textwidth]{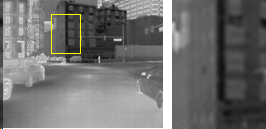}
        \caption{Thermal HR}
        \label{fig:gull}
    \end{subfigure}
    \caption{Visual comparison on an image from FLIR-ADAS test set for $\times$2 SR.}
    \label{fig:x2}
\end{figure*}
\begin{figure*}[hbt!]
    \centering
    \begin{subfigure}[t]{0.195\textwidth}
       \includegraphics[width=\textwidth]{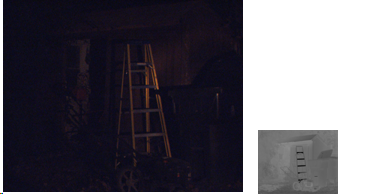}
        \caption{RGB/ Thermal LR}
        \label{fig:gull}
    \end{subfigure}
    \hfill
    \begin{subfigure}[t]{0.195\textwidth}
       \includegraphics[width=\textwidth]{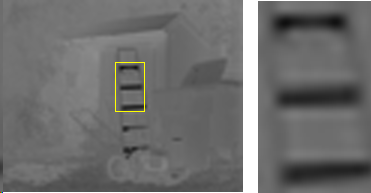}
        \caption{Bicubic}
        \label{fig:gull}
    \end{subfigure}
    \hfill
    \begin{subfigure}[t]{0.195\textwidth}
       \includegraphics[width=\textwidth]{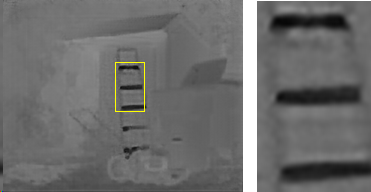}
        \caption{SAN \cite{san}}
        \label{fig:gull}
    \end{subfigure}
    \hfill
    \begin{subfigure}[t]{0.195\textwidth}
       \includegraphics[width=\textwidth]{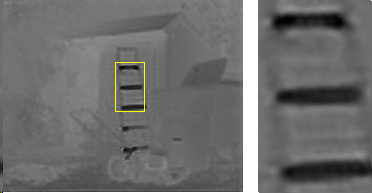}
        \caption{ShuffleMixer \cite{shuffleMixer}}
        \label{fig:gull}
    \end{subfigure}
    \hfill
    \begin{subfigure}[t]{0.195\textwidth}
       \includegraphics[width=\textwidth]{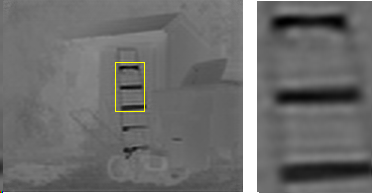}
        \caption{LBNet \cite{lbnet}}
        \label{fig:gull}
    \end{subfigure}
    \hfill
    \begin{subfigure}[t]{0.195\textwidth}
       \includegraphics[width=\textwidth]{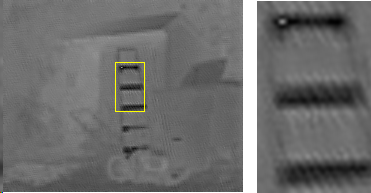}
        \caption{MMSR \cite{mmsr}}
        \label{fig:gull}
    \end{subfigure}
    \hfill
    \begin{subfigure}[t]{0.195\textwidth}
       \includegraphics[width=\textwidth]{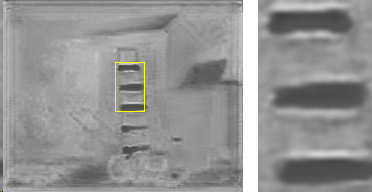}
        \caption{UGSR-FA \cite{ugsr}}
        \label{fig:gull}
    \end{subfigure}
    \hfill
    \begin{subfigure}[t]{0.195\textwidth}
       \includegraphics[width=\textwidth]{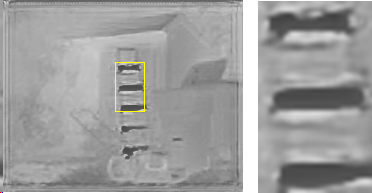}
        \caption{UGSR-ME \cite{ugsr}}
        \label{fig:gull}
    \end{subfigure}
    \hfill
    \begin{subfigure}[t]{0.195\textwidth}
       \includegraphics[width=\textwidth]{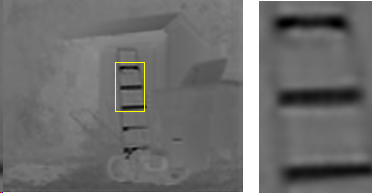}
        \caption{ LATIS (\textbf{Ours})}
        \label{fig:gull}
    \end{subfigure}
    \hfill
    \begin{subfigure}[t]{0.195\textwidth}
       \includegraphics[width=\textwidth]{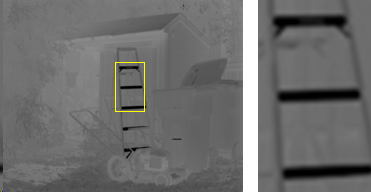}
        \caption{Thermal HR}
        \label{fig:gull}
    \end{subfigure}
    \caption{Visual comparison on an image from CATS test set for $\times$3 SR.}
    \label{fig:x3}
\end{figure*}
\subsubsection{Patchwise EMD loss}
\label{sec:emd}
In our patchwise EMD loss, we formulate a differentiable loss function that compares the super-resolved image $I_{SR}$ and the ground truth $I_{HR}$ based on patchwise histogram similarity. We adopt patchwise comparison considering the fact that the patches at the same spatial locations in $I_{SR}$ and $I_{HR}$ should have similar color distribution. The Earth Mover's Distance (EMD) is used to compare the histograms. 

First, we split $I_{SR}$ and $I_{HR}$ into similar non-overlapping patches, where we consider that in a single-channel thermal image, each patch has a discrete range of 256 intensity values. For any image patch $I_P(x)\in [0,1]$, the $k$-th intensity value is $\frac{k}{255}$. Now, the color distribution of $I_P$ is described by its histogram, which basically counts the number of pixels in each intensity value. For the $k$-th bin, the histogram value is the number of pixels with intensity value $\frac{k}{255}$. However, this counting operation is not differentiable, which makes histogram operation non-differentiable. Hence, we approximate this counting operation with a differentiable function to construct a differentiable histogram. 

We use a series of smooth and differentiable activation functions to approximate the counting operation in the histogram. The idea is that, for counting the number of pixels with $k$-th intensity value $\big(=\frac{k}{255}\big)$, the $k$-th activation function $\Pi_k$ is applied on the image patch $I_P$. This generates an activation map $A_{I_P}(k)$ which approximates a binary response, such that only pixels in $I_P$ with intensity value closer to $\frac{k}{255}$ have higher values in $A_{I_P}(k)$. Thus, the sum of pixels in $A_{I_P}(k)$ approximates the number of pixels in $I_P$ with intensity value $\frac{k}{255}$. This, in turn, is the histogram value in the $k$-th bin.

 To realize such activation functions, first, we partition the pixel intensity interval $[0,1]$ into 256 subintervals $\{B_k\}_{k=0}^{255}$. These subintervals basically approximate the 256 subintervals of the histogram. The $k$-th subinterval $B_k$ has a length $L=\frac{1}{256}$ and center $\mu _k = -1+L(k+\frac{1}{2})$, so $B_k$ can be represented as $B_k=[-1+kL,-1+(k+1)L]$. Here, $\mu _k$ is the same as the $k$-th bin of the histogram. 
 
 Then, we use a differentiable approximation of the rectangular function as the k-th activation function $\Pi_k$, which is defined as,
\begin{equation}
    \Pi_k(y)\triangleq\sigma\left(\frac{y-\mu_k+L/2}{W}\right)-\sigma\left(\frac{y-\mu_k-L/2}{W}\right)
\end{equation}
where $\sigma$ is the softmax function, $W$ is the bandwidth of k-th activation function. Now, applying $\Pi_k$ on image patch $I_P$ thresholds it in the subinterval $B_k$, and generates the $k^{th}$ activation map $A_{I_P}(k)$. 
\begin{equation}
    A_{I_P}(k)=\Pi_k(I_P(x))
\end{equation}
Figure \ref{fig:hist} illustrates the idea of constructing a differentiable histogram using activation functions. Activation functions $\Pi_{130}$, $\Pi_{150}$, $\Pi_{170}$ are applied on an image patch $I_P$ and generate activation maps $A_{I_P}(130)$, $A_{I_P}(150)$, $A_{I_P}(170)$, respectively. The sum of the pixel values in $A_{I_P}(k)$, approximates the value of the histogram in the $k^{th}$ bin. With 256 values of $k$, we get 256 activation maps, and their summation approximates the histogram values in 256 bins. Differentiable normalized histogram $h$ is denoted as,
\begin{equation}
    h=\left\{\mu_k,\frac{1}{N}\sum_{I_P(x)\in [0,1]} A_{I_P}(k)\right\}_{k=0}^{255}
\end{equation}
where $N$ is the number of pixels in $I_P$. Let $h_1$ and $h_2$ be the patchwise normalized histograms of $I_{SR}$ and $I_{HR}$, respectively. To compare between $h_1$ and $h_2$, we use EMD, which equals the $L_1$ distance between the cumulative histograms \cite{emd_l1}. Similar to \cite{emd_l2}, we use the mean squared error (MSE) loss between the cumulative histograms for faster convergence. Patch-wise EMD loss $\mathcal{L}_P$ is formulated as,
\begin{equation}
    \mathcal{L}_P = \frac{1}{M}\big|\big|\:CDF(h_1)-CDF(h_2)\:\big|\big|_2^2
\label{eq:patch}
\end{equation}
where $M$ is the total number of image pixels, $||\cdot||_2$ is the $L_2$ norm, and $CDF(\cdot)$ is the cumulative density function.

\begin{table*}
\centering
\caption{Comparisons on 3 test sets for SR networks. The FLOPs are computed under the setting of super-resolving an LR image of resolution 80$\times$64. The best performances are highlighted in \textcolor{red}{\textbf{red}} and second-best performances are underlined in \textcolor{blue}{\underline{blue}} color.}
\begin{tabular}{LLLLLCCC}
\toprule
\multicolumn{1}{l}{Scale} & \multicolumn{1}{l}{Method} & \multicolumn{1}{l}{\makecell{Single/ \\ Guided}} & \multicolumn{1}{l}{Params} & \multicolumn{1}{l}{FLOPs}& \multicolumn{1}{c}{\makecell{FLIR-ADAS \\ PSNR/SSIM}}
& \multicolumn{1}{c}{\makecell{KAIST \\ PSNR/SSIM}} & \multicolumn{1}{c}{\makecell{CATS \\ PSNR/SSIM}}\\
\midrule
\multirow{ 9}{*}{$\times$2} & \multirow{1}{*}{Bicubic} & \multirow{1}{*}{Single} & - & - & 31.35/0.9047 & 40.27/0.9677 & 36.46/0.9452\\
& \multirow{1}{*}{SAN \cite{san}} & \multirow{1}{*}{Single} & 15297\text{K} & 78.24\text{G} & \textcolor{blue}{\underline{32.01}}/\textcolor{blue}{\underline{0.9182}} & 41.71/0.9688& \textcolor{blue}{\underline{37.35}}/0.9515\\
& \multirow{1}{*}{ShuffleMixer \cite{shuffleMixer}} & \multirow{1}{*}{Single} & 392\text{K} & 1.99\text{G} &  31.77/0.9145 & \textcolor{blue}{\underline{41.94}}/0.9700& 37.12/0.9528\\
& \multirow{1}{*}{LBNet \cite{lbnet}} & \multirow{1}{*}{Single} & 728\text{K} & 4.19\text{G} &  31.84/0.9164 & 41.92/\textcolor{blue}{\underline{0.9706}}& 36.98/\textcolor{blue}{\underline{0.9536}}\\
& \multirow{1}{*}{DCT-Net \cite{dctnet}} & \multirow{1}{*}{Guided} & 483\text{K} & 0.59\text{G} &  10.84/0.3876
 & 20.19/0.5625
& 10.21/0.3747
\\
& \multirow{1}{*}{MMSR \cite{mmsr}} & \multirow{1}{*}{Guided} & 249\text{K} & 5.09\text{G} &  29.45/0.8359
 & 37.54/0.8940
& 31.48/0.8684
\\
& \multirow{1}{*}{UGSR-FA \cite{ugsr}} & \multirow{1}{*}{Guided} & 512\text{K} & 11.51\text{G} &  25.48/0.8557
 & 38.67/0.9548
& 24.90/0.8259
\\
& \multirow{1}{*}{UGSR-ME \cite{ugsr}} & \multirow{1}{*}{Guided} & 2865\text{K} & 38.19\text{G} &  25.83/0.8608
 & 39.46/0.9568
& 24.79/0.8322
\\
& \multirow{1}{*}{LATIS (Ours)} & \multirow{1}{*}{Single} & 193\text{K} & 0.37\text{G} &  \textcolor{red}{\textbf{32.23}}/\textcolor{red}{\textbf{0.9226}} & \textcolor{red}{\textbf{41.98}}/\textcolor{red}{\textbf{0.9717}}& \textcolor{red}{\textbf{37.91}}/\textcolor{red}{\textbf{0.9596}}\\
\midrule
\multirow{ 9}{*}{$\times$3} & \multirow{1}{*}{Bicubic} & \multirow{1}{*}{Single} & - & - &  29.53/0.8386 & 38.64/0.9536 & 34.89/0.9150\\
& \multirow{1}{*}{SAN \cite{san}} & \multirow{1}{*}{Single} & 15482\text{K} & 79.20\text{G} & 29.72/0.8462 & 40.06/0.9516& 35.07/0.9148\\
& \multirow{1}{*}{ShuffleMixer \cite{shuffleMixer}} & \multirow{1}{*}{Single} & 412\text{K} & 2.11\text{G} &  29.61/0.8463 & \textcolor{red}{\textbf{40.46}}/0.9550& 34.95/0.9221\\
& \multirow{1}{*}{LBNet \cite{lbnet}} & \multirow{1}{*}{Single} & 730\text{K} & 4.20\text{G} &  \textcolor{blue}{\underline{29.91}}/\textcolor{blue}{\underline{0.8506}} & \textcolor{blue}{\underline{40.39}}/\textcolor{blue}{\underline{0.9553}}& \textcolor{blue}{\underline{35.39}}/\textcolor{blue}{\underline{0.9229}}\\
& \multirow{1}{*}{DCT-Net \cite{dctnet}} & \multirow{1}{*}{Guided} & 483\text{K} & 1.32\text{G} &  9.30/0.4440
 & 20.20/0.5982
& 10.17/0.4458
\\
& \multirow{1}{*}{MMSR \cite{mmsr}} & \multirow{1}{*}{Guided} & 249\text{K} & 11.44\text{G} &  28.25/0.7731
 & 36.12/0.9018
& 31.98/0.8434
\\
& \multirow{1}{*}{UGSR-FA \cite{ugsr}} & \multirow{1}{*}{Guided} & 512\text{K} & 19.23\text{G} &  24.31/0.7928
 & 37.93/0.9403
& 24.16/0.8000
\\
& \multirow{1}{*}{UGSR-ME \cite{ugsr}} & \multirow{1}{*}{Guided} & 2865\text{K} & 50.17\text{G} &  25.21/0.7938
 & 38.34/0.9404
& 24.38/0.8068
\\
& \multirow{1}{*}{LATIS (Ours)} & \multirow{1}{*}{Single} & 198\text{K} & 0.41\text{G} &  \textcolor{red}{\textbf{30.06}}/\textcolor{red}{\textbf{0.8568}}
 & 40.35/\textcolor{red}{\textbf{0.9558}}
& \textcolor{red}{\textbf{35.81}}/\textcolor{red}{\textbf{0.9298}}
\\
\midrule
\multirow{ 9}{*}{$\times$4} & \multirow{1}{*}{Bicubic} & \multirow{1}{*}{Single} & - & - &  28.43/0.7754 & 37.97/0.9459 & 33.90/0.8950\\
& \multirow{1}{*}{SAN \cite{san}} & \multirow{1}{*}{Single} & 15445\text{K} & 81.30\text{G} & 28.53/0.7791 & 39.56/0.9445& 34.05/0.8917\\
& \multirow{1}{*}{ShuffleMixer \cite{shuffleMixer}} & \multirow{1}{*}{Single} & 408\text{K} & 2.36\text{G} & 28.42/\textcolor{blue}{\underline{0.7835}} & \textcolor{blue}{\underline{39.76}}/\textcolor{blue}{\underline{0.9484}}& 33.64/\textcolor{blue}{\underline{0.9024}}\\
& \multirow{1}{*}{LBNet \cite{lbnet}} & \multirow{1}{*}{Single} & 732\text{K} & 4.21\text{G} &  \textcolor{blue}{\underline{28.67}}/0.7829 & 39.75/0.9468& \textcolor{blue}{\underline{34.22}}/0.9003\\
& \multirow{1}{*}{DCT-Net \cite{dctnet}} & \multirow{1}{*}{Guided} & 483\text{K} & 2.35\text{G} &  10.20/0.4636
 & 20.79/0.6154
& 10.26/0.4353
\\
& \multirow{1}{*}{MMSR \cite{mmsr}} & \multirow{1}{*}{Guided} & 249\text{K} & 20.34\text{G} &  27.24/0.6995 & 35.68/0.8947& 31.49/0.8173\\
& \multirow{1}{*}{UGSR-FA \cite{ugsr}} & \multirow{1}{*}{Guided} & 1150\text{K} & 47.55\text{G} &  24.23/0.7230
 & 38.06/0.9356
& 24.12/0.7588
\\
& \multirow{1}{*}{UGSR-ME \cite{ugsr}} & \multirow{1}{*}{Guided} & 3679\text{K} & 90.23\text{G} &  24.57/0.7396
 & 38.04/0.9316
& 24.59/0.7870
\\
& \multirow{1}{*}{LATIS (Ours)} & \multirow{1}{*}{Single} & 197\text{K} & 0.47\text{G} &  \textcolor{red}{\textbf{28.85}}/\textcolor{red}{\textbf{0.7916}}
 & \textcolor{red}{\textbf{39.81}}/\textcolor{red}{\textbf{0.9484}}
& \textcolor{red}{\textbf{34.69}}/\textcolor{red}{\textbf{0.9099}}
\\
\bottomrule
\end{tabular}
\label{tab:example}
\end{table*}
\section{Experiments}
\label{sec:experiments}
\subsection{Setup}
\label{sec:setup}
\textbf{Implementation details.} Our model, LATIS, consists of three LGFBs. In the CSConv layer, we use two convolution layers with kernel sizes of $3\times 3$ and $7\times 7$ sequentially. In the channel shuffling operation of the CSConv layer, we use a group size of 4. In the linear projection layer of GFE module, we set the query and key depth to 16, the value depth to 8, and the number of heads to 4. We use the 3D convolution operation to perform relative positional encoding in the lambda abstraction layer of the GFE module. The kernel size of 3D convolution is set to $1\times 25\times 25$. In LATIS, the intermediate feature channel dimension is 32. 

For our proposed patch-wise EMD loss $\mathcal{L}_P$, we use a patch size of $8\times 8$. In equation \ref{eq:lossfn}, the number of epochs $n$ is set to 5.
\\
\textbf{Datasets.} We select the first 1500 thermal-RGB image pairs of the FLIR-ADAS \cite{flir_adas} dataset along with the first 500 image pairs of the KAIST \cite{kaist} dataset as the train set (a total of 2000 image pairs). For SISR, only the thermal images are employed. We utilize 375 image pairs from FLIR-ADAS, 101 image pairs from KAIST, and 67 image pairs from the CATS \cite{cats} dataset as test sets. We train the models on the train set and test on these three test sets mentioned above. In our experiments, LR thermal images of size $80\times 64$ are generated from corresponding HR thermal images using the bicubic degradation method. More details about the datasets are given in the supplementary material. 
\\
\textbf{Training settings.} Our model,LATIS is trained by minimizing $\mathcal{L} = \mathcal{L}_C + \lambda \mathcal{L}_P$ loss, as in equation \ref{eq:lossfn} with Adam \cite{adam} optimizer for 200 epochs. We employed a batch size of 64 for $\times$2 scale, 48 for $\times$3 scale, and 32 for $\times$4 scale. We keep the learning rate constant at  $1 \times 10^{-4}$. The LATIS model is implemented with the PyTorch framework using an Nvidia A40 GPU. 
\\
\textbf{Metrics.} We conduct experiments, for the upscaling factors of $\times$2, $\times$3, and $\times$4. During the testing stage, we compare the SR performance based on peak signal-to-noise ratio (PSNR) and structural similarity index (SSIM) \cite{ssim}.

\subsection{Comparison with state-of-the-art methods}
To evaluate the performance, we compare LATIS with both GSR and SISR methods proposed for visible, depth, or thermal images. Among the SISR methods, we compare with four methods: Bicubic, SAN \cite{san}, ShuffleMixer \cite{shuffleMixer}, and LBNet \cite{lbnet}. Among the GSR methods, we compare with four methods: DCTNet \cite{dctnet}, MMSR \cite{mmsr}, UGSR-FA \cite{ugsr}, and UGSR-ME \cite{ugsr}. We train all the models with our training dataset before testing.  
\\
\textbf{Visual Comparison.} For visual comparison, Figure \ref{fig:x2} presents an image from the FLIR-ADAS test set for $\times$2 scale. Similarly, figure \ref{fig:x3} presents another image from the CATS test set for $\times$3 scale. From the visual representation, it is clear that the overall reconstruction ability of LATIS is better in comparison to the other methods. To make it visually more evident, we select a patch in the image, depicted using a yellow box, and zoom into it. As shown in the zoomed versions, LATIS reconstructs the building edge and the ladder steps with comparably more structural similarity and fewer artifacts in the background. More visual comparisons are shown in the supplementary material. 
\\
\textbf{Quantitative Comparison.} Table \ref{tab:example} presents the quantitative results on three test sets with scaling factors of $\times$2, $\times$3, and $\times$4. Along with PSNR and SSIM metrics, we list the model parameters and FLOPs. We calculate the FLOPs under the setting of super-resolving an LR image of resolution $80\times 64$. 

As the results show, SISR methods perform better than the guided methods. DCTNet, a GSR method for depth images, performs poorly in all the test sets. As mentioned in the original paper \cite{dctnet}, the performance of DCTNet is based on the assumption that the occurrence of texture edges in RGB image and the edge discontinuities in corresponding depth image are statistically correlated. Maybe this assumption does not hold good for thermal-visible image-pairs.

In comparison to other methods, LATIS achieves competitive performance with the lowest number of model parameters and FLOPs. Especially, LATIS has a comparable number of parameters to MMSR \cite{mmsr}, but our model outperforms it by a large margin on all the test sets.
LATIS performs better than other methods, except for PSNR value on the KAIST test set for $\times 3$ scale, compared to ShuffleMixer \cite{shuffleMixer} and LBNet \cite{lbnet}. LATIS achieves significant gain on the CATS test set for $\times 3$ scale, yielding a 0.55dB improvement in the PSNR value. All these results demonstrate the effectiveness of our proposed method.

\subsection{Ablation Study}
\label{sec:ablation}
To verify the effectiveness of our proposed modules and loss function, we conduct an ablation study. For the ablation experiments, we train the models with our training dataset described in section \ref{sec:setup}. For model ablations, we test on the CATS test set for $\times$2 scale. FLOPs are computed under the setting of super-resolving an LR image of resolution 80$\times$64.

\noindent
\textbf{Effects of changing number of LGFB in LATIS: }
In LATIS, we use three LGFBs. To see the effect of different numbers of LGFB, we conduct experiments for three values: 2, 3, and 4. Table \ref{tab:lgfb} shows the results. LATIS has the best performance when three LGFBs are used.
\begin{table}[htb]
\caption{Effects of changing number of  LGFBs in LATIS.}
\label{tab:lgfb}
\begin{tabularx}{0.48\textwidth} 
{ 
   >{\centering\arraybackslash}X 
  | >{\centering\arraybackslash}X 
  >{\centering\arraybackslash}X
  >{\centering\arraybackslash}X}
 \hline
 Number & 2 & 3 & 4\\
 \hline
 PSNR/SSIM & 37.34/0.9562 & \textbf{37.91}/\textbf{0.9596}& 37.65/0.9583\\
 \hline
\end{tabularx}
\end{table}

\noindent
\textbf{Effectiveness of lambda abstraction in GFE module: }
In the GFE module, we use the lambda abstraction to model long-range dependency. We conduct experiments to verify the effectiveness of the lambda abstraction with respect to the self-attention mechanism. We build an alternate model by replacing the lambda abstraction layer in the GFE module with a multi-head self-attention \cite{vit} layer. In the self-attention layer, the feature channel dimension is 32, the window size is 16$\times$16, and the number of heads is 4. More details about the self-attention module are given in the supplementary material. Table \ref{tab:vit} presents the comparison results. Though the self-attention based model has fewer parameters, using the lambda abstraction-based GFE module has lesser model complexity and improves the PSNR and SSIM values by a large margin. The PSNR value is improved by 1.15 dB and the SSIM value by 0.0107. 
\begin{table}[hbt!]
\caption{Effectiveness of the lambda abstraction in GFE module.}
\centering
\begin{tabular}{CCCC}
\toprule
\multicolumn{1}{c}{Structure} & \multicolumn{1}{c}{w/ Self-attention}& \multicolumn{1}{c}{w/ Lambda abstraction} \\
\midrule
\multicolumn{1}{c}{PSNR/SSIM} & \multicolumn{1}{c}{36.76/0.9489}& \multicolumn{1}{c}{\textbf{37.91}/\textbf{0.9596}} \\
\multicolumn{1}{c}{Params (K)} & \multicolumn{1}{c}{86
}& \multicolumn{1}{c}{193} \\
\multicolumn{1}{c}{FLOPs (G)}& \multicolumn{1}{c}{0.43
}& \multicolumn{1}{c}{0.37} \\
\bottomrule
\end{tabular}
\label{tab:vit}
\end{table}

\noindent
\textbf{Effects of layer normalization in GFE module: }
In GFE module, we apply layer normalization before the linear projection. As shown in Table \ref{tab:layerNorm}, layer normalization improves the PSNR value by 2.08 dB and the SSIM value by 0.0212.

\noindent
\textbf{Effects of channel shuffling in CSConv layer: }
In the CSConv layer, we apply channel shuffling operation for better feature representation. As shown in 
Table \ref{tab:channel}, channel shuffling improves the PSNR value by 0.16 dB and the SSIM value by 0.0016.

\noindent
\textbf{Effects of CBAM in LGFB: }
In LGFB, we employ CBAM for adaptive feature refinement. As shown in Table \ref{tab:cbam}, CBAM improves the PSNR value by 0.14 dB and the SSIM value by 0.0007. 

\begin{table}[htb]
\caption{Effects of the layer normalization in GFE module.}
\label{tab:layerNorm}
\begin{tabularx}{0.48\textwidth} 
{ 
   >{\centering\arraybackslash}X 
  | >{\centering\arraybackslash}X 
  >{\centering\arraybackslash}X  }
 \hline
 Structure & w/o layer norm & w/ layer norm\\
 \hline
 PSNR/SSIM & 35.83/0.9384 & \textbf{37.91}/\textbf{0.9596}\\
 \hline
\end{tabularx}
\end{table}
\vspace{-1.5em}
\begin{table}[htb]
\caption{Effects of the channel shuffling (CS) in CSConv layer.}
\label{tab:channel}
\begin{tabularx}{0.48\textwidth} 
{ 
   >{\centering\arraybackslash}X 
  | >{\centering\arraybackslash}X 
  >{\centering\arraybackslash}X  }
 \hline
 Structure & w/o CS & w/ CS\\
 \hline
 PSNR/SSIM & 37.75/0.9580 & \textbf{37.91}/\textbf{0.9596}\\
 \hline
\end{tabularx}
\end{table}
\vspace{-1.5em}
\begin{table}[htb]
\caption{Effects of the CBAM in LGFB.}
\label{tab:cbam}
\begin{tabularx}{0.48\textwidth} 
{ 
   >{\centering\arraybackslash}X 
  | >{\centering\arraybackslash}X 
  >{\centering\arraybackslash}X  }
 \hline
 Structure & w/o CBAM & w/ CBAM\\
 \hline
 PSNR/SSIM & 37.77/0.9589 & \textbf{37.91}/\textbf{0.9596}\\
 \hline
\end{tabularx}
\end{table}
\noindent
\textbf{Effects of patch-wise EMD loss: }
To verify the effectiveness of $\mathcal{L}_P$ loss, once we train our LATIS with only $\mathcal{L}_C$\: loss and another time we train it with $\mathcal{L}_C+\lambda\mathcal{L}_P$\:loss. Results in table \ref{tab:loss} shows that except for the KAIST test set at $\times$3 scale, our proposed $\mathcal{L}_P$ loss improves the PSNR value by 0.02 dB to 0.68 dB and the SSIM value by 0.0003 to 0.0228.
\begin{table}[hbt!]
\centering
\caption{Effectiveness of patchwise EMD loss.}
\begin{tabular}{LCCC}
\toprule
\multicolumn{1}{c}{Dataset} & \multicolumn{1}{c}{Scale} & \multicolumn{1}{c}{\makecell{Training with \\ $\mathcal{L}_C$\:loss \\ PSNR/SSIM}} & \multicolumn{1}{c}{\makecell{Training with\\ 
$\mathcal{L}_C+\lambda\mathcal{L}_P$\:loss \\ PSNR/SSIM}} \\
\midrule
\multirow{ 3}{*}{FLIR-ADAS} & \multirow{1}{*}{$\times$2} & 32.16/0.9212
 & \textbf{32.23}/\textbf{0.9226}
\\
& \multirow{1}{*}{$\times$3} & 30.00/0.8553
 & \textbf{30.06}/\textbf{0.8568}
\\
& \multirow{1}{*}{$\times$4} & 28.76/0.7867
 & \textbf{28.85}/\textbf{0.7916}\\
 \midrule
\multirow{ 3}{*}{KAIST} & \multirow{1}{*}{$\times$2} & 41.94/0.9713
 & \textbf{41.98}/\textbf{0.9717}
\\
& \multirow{1}{*}{$\times$3} & \textbf{40.38}/\textbf{0.9563}
 & 40.35/0.9558
\\
& \multirow{1}{*}{$\times$4} & 39.75/0.9471
 & \textbf{39.81}/\textbf{0.9484}\\
\midrule
\multirow{ 3}{*}{CATS} & \multirow{1}{*}{$\times$2} & 37.55/0.9586
 & \textbf{37.91}/\textbf{0.9596}
\\
& \multirow{1}{*}{$\times$3} & 35.79/0.9295
 & \textbf{35.81}/\textbf{0.9298}
\\
& \multirow{1}{*}{$\times$4} & 34.01/0.8871
 & \textbf{34.69}/\textbf{0.9099}\\
\bottomrule
\end{tabular}
\label{tab:loss}
\end{table}


\section{Conclusion}
\label{sec:conclusion}
We propose Lambda Abstraction-based Thermal Image Super-resolution, LATIS, a lightweight SISR network for thermal images. Our model uses the core module LGFB, to sequentially capture the local and global features. This allows LATIS to encode information efficiently for better SR. Besides, we introduce a global feature extraction module based on the lambda abstraction mechanism and a CSConv layer to explore the local context. Additionally, we propose a patchwise EMD loss to further improve the SR quality. Experimental results demonstrate the effectiveness of our proposed modules and loss function. LATIS achieves competitive performance with state-of-the-art methods while maintaining the least model parameters and complexity. 
\clearpage
{
    \small
    \bibliographystyle{ieeenat_fullname}
    \bibliography{main}

\begin{thebibliography}{47}
\providecommand{\natexlab}[1]{#1}
\providecommand{\url}[1]{\texttt{#1}}
\expandafter\ifx\csname urlstyle\endcsname\relax
  \providecommand{\doi}[1]{doi: #1}\else
  \providecommand{\doi}{doi: \begingroup \urlstyle{rm}\Url}\fi

\bibitem[Avi-Aharon et~al.(2023)Avi-Aharon, Arbelle, and Raviv]{huenet}
Mor Avi-Aharon, Assaf Arbelle, and Tammy~Riklin Raviv.
\newblock Differentiable histogram loss functions for intensity-based image-to-image translation.
\newblock \emph{IEEE Transactions on Pattern Analysis and Machine Intelligence}, 45\penalty0 (10):\penalty0 11642--11653, 2023.

\bibitem[Bello(2021)]{lambdanetworks}
Irwan Bello.
\newblock Lambdanetworks: Modeling long-range interactions without attention.
\newblock In \emph{International Conference on Learning Representations}, 2021.

\bibitem[Brzezinski et~al.(2021)Brzezinski, Rabin, Lewis, Peled, Kerpel, Tsur, Gendelman, Naftali-Shani, Gringauz, Amital, Leibowitz, Mayan, Ben-Zvi, Heller, Shechtman, Rogowski, Shenhar-Tsarfaty, Konen, Marom, Ironi, Rahav, Zimmer, Grossman, Ovadia-Blechman, Leor, and Hoffer]{disease2}
Rafael~Y. Brzezinski, Neta Rabin, Nir Lewis, Racheli Peled, Ariel Kerpel, Avishai~M. Tsur, Omer Gendelman, Nili Naftali-Shani, Irina Gringauz, Howard Amital, Avshalom Leibowitz, Haim Mayan, Ilan Ben-Zvi, Eyal Heller, Liran Shechtman, Ori Rogowski, Shani Shenhar-Tsarfaty, Eli Konen, Edith~M. Marom, Avinoah Ironi, Galia Rahav, Yair Zimmer, Ehud Grossman, Zehava Ovadia-Blechman, Jonathan Leor, and Oshrit Hoffer.
\newblock Automated processing of thermal imaging to detect covid-19.
\newblock \emph{Scientific Reports}, 11\penalty0 (1):\penalty0 17489, 2021.

\bibitem[Chen et~al.(2023{\natexlab{a}})Chen, Wang, Zhou, Qiao, and Dong]{hat}
Xiangyu Chen, Xintao Wang, Jiantao Zhou, Yu Qiao, and Chao Dong.
\newblock Activating more pixels in image super-resolution transformer.
\newblock In \emph{Proceedings of the IEEE/CVF Conference on Computer Vision and Pattern Recognition}, pages 22367--22377, 2023{\natexlab{a}}.

\bibitem[Chen et~al.(2023{\natexlab{b}})Chen, Zhang, Gu, Kong, Yang, and Yu]{dat}
Zheng Chen, Yulun Zhang, Jinjin Gu, Linghe Kong, Xiaokang Yang, and Fisher Yu.
\newblock Dual aggregation transformer for image super-resolution.
\newblock In \emph{ICCV}, 2023{\natexlab{b}}.

\bibitem[Choromanski et~al.(2021)Choromanski, Likhosherstov, Dohan, Song, Gane, Sarlos, Hawkins, Davis, Mohiuddin, Kaiser, Belanger, Colwell, and Weller]{performer}
Krzysztof~Marcin Choromanski, Valerii Likhosherstov, David Dohan, Xingyou Song, Andreea Gane, Tamas Sarlos, Peter Hawkins, Jared~Quincy Davis, Afroz Mohiuddin, Lukasz Kaiser, David~Benjamin Belanger, Lucy~J Colwell, and Adrian Weller.
\newblock Rethinking attention with performers.
\newblock In \emph{International Conference on Learning Representations}, 2021.

\bibitem[Dai et~al.(2019)Dai, Cai, Zhang, Xia, and Zhang]{san}
Tao Dai, Jianrui Cai, Yongbing Zhang, Shu-Tao Xia, and Lei Zhang.
\newblock Second-order attention network for single image super-resolution.
\newblock In \emph{2019 IEEE/CVF Conference on Computer Vision and Pattern Recognition (CVPR)}, pages 11057--11066, 2019.

\bibitem[Deng and Dragotti(2020)]{ista}
Xin Deng and Pier~Luigi Dragotti.
\newblock Deep coupled ista network for multi-modal image super-resolution.
\newblock \emph{IEEE Transactions on Image Processing}, 29:\penalty0 1683--1698, 2020.

\bibitem[Dong et~al.(2014)Dong, Loy, He, and Tang]{srcnn}
Chao Dong, Chen~Change Loy, Kaiming He, and Xiaoou Tang.
\newblock Learning a deep convolutional network for image super-resolution.
\newblock In \emph{Computer Vision -- ECCV 2014}, pages 184--199, Cham, 2014. Springer International Publishing.

\bibitem[Dong et~al.(2016)Dong, Loy, and Tang]{fsrcnn}
Chao Dong, Chen~Change Loy, and Xiaoou Tang.
\newblock Accelerating the super-resolution convolutional neural network.
\newblock In \emph{Computer Vision--ECCV 2016: 14th European Conference, Amsterdam, The Netherlands, October 11-14, 2016, Proceedings, Part II 14}, pages 391--407. Springer, 2016.

\bibitem[Dong et~al.(2022)Dong, Yokoya, Wang, and Uezato]{mmsr}
Xiaoyu Dong, Naoto Yokoya, Longguang Wang, and Tatsumi Uezato.
\newblock Learning mutual modulation for self-supervised cross-modal super-resolution.
\newblock In \emph{Computer Vision -- ECCV 2022}, pages 1--18, Cham, 2022. Springer Nature Switzerland.

\bibitem[Dosovitskiy et~al.(2021)Dosovitskiy, Beyer, Kolesnikov, Weissenborn, Zhai, Unterthiner, Dehghani, Minderer, Heigold, Gelly, Uszkoreit, and Houlsby]{vit}
Alexey Dosovitskiy, Lucas Beyer, Alexander Kolesnikov, Dirk Weissenborn, Xiaohua Zhai, Thomas Unterthiner, Mostafa Dehghani, Matthias Minderer, Georg Heigold, Sylvain Gelly, Jakob Uszkoreit, and Neil Houlsby.
\newblock An image is worth 16x16 words: Transformers for image recognition at scale.
\newblock In \emph{International Conference on Learning Representations}, 2021.

\bibitem[FLIR(2022)]{flir_adas}
Teledyne FLIR.
\newblock Teledyne flir free adas thermal dataset v2.
\newblock \url{https://adas-dataset-v2.flirconservator.com/\#downloadguide}, 2022.
\newblock Accessed on: 2023-01-07.

\bibitem[Gade and Moeslund(2013)]{survei}
Rikke Gade and Thomas~Baltzer Moeslund.
\newblock Thermal cameras and applications: a survey.
\newblock \emph{Machine Vision and Applications}, 25:\penalty0 245 -- 262, 2013.

\bibitem[Gao et~al.(2022)Gao, Wang, Li, Li, Yu, and Zeng]{lbnet}
Guangwei Gao, Zhengxue Wang, Juncheng Li, Wenjie Li, Yi Yu, and Tieyong Zeng.
\newblock Lightweight bimodal network for single-image super-resolution via symmetric cnn and recursive transformer.
\newblock In \emph{International Joint Conference on Artificial Intelligence}, 2022.

\bibitem[Guo et~al.(2019)Guo, Li, Guo, Cong, Fu, and Han]{depthsrnet}
Chunle Guo, Chongyi Li, Jichang Guo, Runmin Cong, Huazhu Fu, and Ping Han.
\newblock Hierarchical features driven residual learning for depth map super-resolution.
\newblock \emph{IEEE Transactions on Image Processing}, 28\penalty0 (5):\penalty0 2545--2557, 2019.

\bibitem[Gupta and Mitra(2021)]{ugsr}
Honey Gupta and Kaushik Mitra.
\newblock Toward unaligned guided thermal super-resolution.
\newblock \emph{IEEE Transactions on Image Processing}, 31:\penalty0 433--445, 2021.

\bibitem[Gutierrez and Beksi(2021)]{icvs_thermal}
Nolan~B Gutierrez and William~J Beksi.
\newblock Thermal image super-resolution using second-order channel attention with varying receptive fields.
\newblock In \emph{Computer Vision Systems: 13th International Conference, ICVS 2021, Virtual Event, September 22-24, 2021, Proceedings 13}, pages 3--13. Springer, 2021.

\bibitem[Hou et~al.(2017)Hou, Yu, and Samaras]{emd_l2}
Le Hou, Chen-Ping Yu, and Dimitris Samaras.
\newblock Squared earth movers distance loss for training deep neural networks on ordered-classes.
\newblock In \emph{NIPS Workshop}, 2017.

\bibitem[Huang et~al.(2022)Huang, Miyazaki, Liu, and Omachi]{infrared}
Yongsong Huang, Tomo Miyazaki, Xiaofeng Liu, and Shinichiro Omachi.
\newblock Infrared image super-resolution: Systematic review, and future trends.
\newblock \emph{arXiv preprint arXiv:2212.12322}, 2022.

\bibitem[Hui et~al.(2016)Hui, Loy, and Tang]{msgnet}
Tak-Wai Hui, Chen~Change Loy, and Xiaoou Tang.
\newblock Depth map super-resolution by deep multi-scale guidance.
\newblock In \emph{Computer Vision--ECCV 2016: 14th European Conference, Amsterdam, The Netherlands, October 11-14, 2016, Proceedings, Part III 14}, pages 353--369. Springer, 2016.

\bibitem[Hwang et~al.(2015)Hwang, Park, Kim, Choi, and So~Kweon]{kaist}
Soonmin Hwang, Jaesik Park, Namil Kim, Yukyung Choi, and In So~Kweon.
\newblock Multispectral pedestrian detection: Benchmark dataset and baseline.
\newblock In \emph{Proceedings of the IEEE conference on computer vision and pattern recognition}, pages 1037--1045, 2015.

\bibitem[Jain et~al.(2023)Jain, Li, Chiu, Hassani, Orlov, and Shi]{oneformer}
Jitesh Jain, Jiachen Li, Mang~Tik Chiu, Ali Hassani, Nikita Orlov, and Humphrey Shi.
\newblock Oneformer: One transformer to rule universal image segmentation.
\newblock In \emph{Proceedings of the IEEE/CVF Conference on Computer Vision and Pattern Recognition}, pages 2989--2998, 2023.

\bibitem[Kim et~al.(2016)Kim, Lee, and Lee]{vdsr}
Jiwon Kim, Jung~Kwon Lee, and Kyoung~Mu Lee.
\newblock Accurate image super-resolution using very deep convolutional networks.
\newblock In \emph{Proceedings of the IEEE conference on computer vision and pattern recognition}, pages 1646--1654, 2016.

\bibitem[Kingma and Ba(2015)]{adam}
Diederik~P. Kingma and Jimmy Ba.
\newblock Adam: A method for stochastic optimization.
\newblock In \emph{ICLR}, 2015.

\bibitem[Lai et~al.(2017)Lai, Huang, Ahuja, and Yang]{lapsrn}
Wei-Sheng Lai, Jia-Bin Huang, Narendra Ahuja, and Ming-Hsuan Yang.
\newblock Deep laplacian pyramid networks for fast and accurate super-resolution.
\newblock In \emph{2017 IEEE Conference on Computer Vision and Pattern Recognition (CVPR)}, pages 5835--5843, 2017.

\bibitem[Ledig et~al.(2017)Ledig, Theis, Husz{\'a}r, Caballero, Cunningham, Acosta, Aitken, Tejani, Totz, Wang, et~al.]{srgan}
Christian Ledig, Lucas Theis, Ferenc Husz{\'a}r, Jose Caballero, Andrew Cunningham, Alejandro Acosta, Andrew Aitken, Alykhan Tejani, Johannes Totz, Zehan Wang, et~al.
\newblock Photo-realistic single image super-resolution using a generative adversarial network.
\newblock In \emph{Proceedings of the IEEE conference on computer vision and pattern recognition}, pages 4681--4690, 2017.

\bibitem[Li et~al.(2020)Li, Zhou, Qi, Jiang, Lu, and Jia]{lapar}
Wenbo Li, Kun Zhou, Lu Qi, Nianjuan Jiang, Jiangbo Lu, and Jiaya Jia.
\newblock Lapar: Linearly-assembled pixel-adaptive regression network for single image super-resolution and beyond.
\newblock \emph{Advances in Neural Information Processing Systems}, 33:\penalty0 20343--20355, 2020.

\bibitem[Liang et~al.(2021)Liang, Cao, Sun, Zhang, Van~Gool, and Timofte]{swinir}
Jingyun Liang, Jiezhang Cao, Guolei Sun, Kai Zhang, Luc Van~Gool, and Radu Timofte.
\newblock Swinir: Image restoration using swin transformer.
\newblock In \emph{Proceedings of the IEEE/CVF international conference on computer vision}, pages 1833--1844, 2021.

\bibitem[Lim et~al.(2017)Lim, Son, Kim, Nah, and Mu~Lee]{edsr}
Bee Lim, Sanghyun Son, Heewon Kim, Seungjun Nah, and Kyoung Mu~Lee.
\newblock Enhanced deep residual networks for single image super-resolution.
\newblock In \emph{Proceedings of the IEEE conference on computer vision and pattern recognition workshops}, pages 136--144, 2017.

\bibitem[Miethig et~al.(2019)Miethig, Liu, Habibi, and Mohrenschildt]{vehi}
Ben Miethig, Ash Liu, Saeid Habibi, and Martin~v. Mohrenschildt.
\newblock Leveraging thermal imaging for autonomous driving.
\newblock In \emph{2019 IEEE Transportation Electrification Conference and Expo (ITEC)}, pages 1--5, 2019.

\bibitem[Pitie et~al.(2005)Pitie, Kokaram, and Dahyot]{probability}
Francois Pitie, Anil~C Kokaram, and Rozenn Dahyot.
\newblock N-dimensional probability density function transfer and its application to color transfer.
\newblock In \emph{Tenth IEEE International Conference on Computer Vision (ICCV'05) Volume 1}, pages 1434--1439. IEEE, 2005.

\bibitem[Reinhard et~al.(2001)Reinhard, Adhikhmin, Gooch, and Shirley]{color}
Erik Reinhard, Michael Adhikhmin, Bruce Gooch, and Peter Shirley.
\newblock Color transfer between images.
\newblock \emph{IEEE Computer graphics and applications}, 21\penalty0 (5):\penalty0 34--41, 2001.

\bibitem[Shi et~al.(2016)Shi, Caballero, Husz{\'a}r, Totz, Aitken, Bishop, Rueckert, and Wang]{pixelshuffle}
Wenzhe Shi, Jose Caballero, Ferenc Husz{\'a}r, Johannes Totz, Andrew~P Aitken, Rob Bishop, Daniel Rueckert, and Zehan Wang.
\newblock Real-time single image and video super-resolution using an efficient sub-pixel convolutional neural network.
\newblock In \emph{Proceedings of the IEEE conference on computer vision and pattern recognition}, pages 1874--1883, 2016.

\bibitem[Sun et~al.(2022)Sun, Pan, and Tang]{shuffleMixer}
Long Sun, Jinshan Pan, and Jinhui Tang.
\newblock {ShuffleMixer}: An efficient convnet for image super-resolution.
\newblock In \emph{Advances in Neural Information Processing Systems}, 2022.

\bibitem[Torra et~al.(2022)Torra, Viela, Megías, Sot, and Flors]{disease1}
Joaquim Torra, Felipe Viela, Diego Megías, Begoña Sot, and Cristina Flors.
\newblock Versatile near-infrared super-resolution imaging of amyloid fibrils with the fluorogenic probe cranad-2.
\newblock \emph{Chemistry – A European Journal}, 28\penalty0 (19):\penalty0 e202200026, 2022.

\bibitem[Treible et~al.(2017)Treible, Saponaro, Sorensen, Kolagunda, O'Neal, Phelan, Sherbondy, and Kambhamettu]{cats}
Wayne Treible, Philip Saponaro, Scott Sorensen, Abhishek Kolagunda, Michael O'Neal, Brian Phelan, Kelly Sherbondy, and Chandra Kambhamettu.
\newblock Cats: A color and thermal stereo benchmark.
\newblock In \emph{Proceedings of the IEEE Conference on Computer Vision and Pattern Recognition}, pages 2961--2969, 2017.

\bibitem[Vaswani et~al.(2017)Vaswani, Shazeer, Parmar, Uszkoreit, Jones, Gomez, Kaiser, and Polosukhin]{attention}
Ashish Vaswani, Noam Shazeer, Niki Parmar, Jakob Uszkoreit, Llion Jones, Aidan~N Gomez, {\L}ukasz Kaiser, and Illia Polosukhin.
\newblock Attention is all you need.
\newblock \emph{Advances in neural information processing systems}, 30, 2017.

\bibitem[Wang et~al.(2020)Wang, Li, Khabsa, Fang, and Ma]{linformer}
Sinong Wang, Belinda~Z Li, Madian Khabsa, Han Fang, and Hao Ma.
\newblock Linformer: Self-attention with linear complexity.
\newblock \emph{arXiv preprint arXiv:2006.04768}, 2020.

\bibitem[Wang et~al.(2018)Wang, Yu, Wu, Gu, Liu, Dong, Qiao, and Change~Loy]{esrgan}
Xintao Wang, Ke Yu, Shixiang Wu, Jinjin Gu, Yihao Liu, Chao Dong, Yu Qiao, and Chen Change~Loy.
\newblock Esrgan: Enhanced super-resolution generative adversarial networks.
\newblock In \emph{Proceedings of the European conference on computer vision (ECCV) workshops}, pages 0--0, 2018.

\bibitem[Wang et~al.(2004)Wang, Bovik, Sheikh, and Simoncelli]{ssim}
Zhou Wang, A.C. Bovik, H.R. Sheikh, and E.P. Simoncelli.
\newblock Image quality assessment: from error visibility to structural similarity.
\newblock \emph{IEEE Transactions on Image Processing}, 13\penalty0 (4):\penalty0 600--612, 2004.

\bibitem[Wang et~al.(2022)Wang, Cun, Bao, Zhou, Liu, and Li]{uformer}
Zhendong Wang, Xiaodong Cun, Jianmin Bao, Wengang Zhou, Jianzhuang Liu, and Houqiang Li.
\newblock Uformer: A general u-shaped transformer for image restoration.
\newblock In \emph{Proceedings of the IEEE/CVF conference on computer vision and pattern recognition}, pages 17683--17693, 2022.

\bibitem[Werman et~al.(1985)Werman, Peleg, and Rosenfeld]{emd_l1}
Michael Werman, Shmuel Peleg, and Azriel Rosenfeld.
\newblock A distance metric for multidimensional histograms.
\newblock \emph{Computer Vision, Graphics, and Image Processing}, 32\penalty0 (3):\penalty0 328--336, 1985.

\bibitem[Woo et~al.(2018)Woo, Park, Lee, and Kweon]{cbam}
Sanghyun Woo, Jongchan Park, Joon-Young Lee, and In~So Kweon.
\newblock Cbam: Convolutional block attention module.
\newblock In \emph{Proceedings of the European conference on computer vision (ECCV)}, pages 3--19, 2018.

\bibitem[Xu et~al.(2019)Xu, Sun, Zhang, Zhao, and Lin]{layerNorm}
Jingjing Xu, Xu Sun, Zhiyuan Zhang, Guangxiang Zhao, and Junyang Lin.
\newblock Understanding and improving layer normalization.
\newblock \emph{Advances in Neural Information Processing Systems}, 32, 2019.

\bibitem[Zhang et~al.(2018)Zhang, Tian, Kong, Zhong, and Fu]{rdn}
Yulun Zhang, Yapeng Tian, Yu Kong, Bineng Zhong, and Yun Fu.
\newblock Residual dense network for image super-resolution.
\newblock In \emph{Proceedings of the IEEE conference on computer vision and pattern recognition}, pages 2472--2481, 2018.

\bibitem[Zhao et~al.(2022)Zhao, Zhang, Xu, Lin, and Pfister]{dctnet}
Zixiang Zhao, Jiangshe Zhang, Shuang Xu, Zudi Lin, and Hanspeter Pfister.
\newblock Discrete cosine transform network for guided depth map super-resolution.
\newblock In \emph{Proceedings of the IEEE/CVF Conference on Computer Vision and Pattern Recognition}, pages 5697--5707, 2022.

\end{thebibliography}
}


\end{document}